\documentclass[aps,prl,preprint,groupedaddress]{revtex4-1}
\usepackage{graphicx}
\usepackage{dcolumn}
\usepackage{bm}
\usepackage{epstopdf}
\usepackage{color}
\usepackage{setspace}

\begin{document}

\title{Intrinsic features of an ideal glass}

\author{D. Y. Sun$^{1,3}$, C. Shang$^{2}$, Z. P. Liu$^{2}$ and X. G. Gong$^{3,4}$ }
\email{Email address: xggong@fudan.edu.cn}
\affiliation{$^{1}$ Department of Physics, East China Normal University,
Shanghai 200062, China}
\affiliation{$^{2}$Department of Chemistry,Fudan University, Shanghai 200433}
\affiliation{$^{3}$Key Laboratory for Computational Physical Sciences (MOE), State Key Laboratory of Surface Physics, Department of Physics, Fudan University, Shanghai 200433, China}
\affiliation{$^{4}$Collaborative Innovation Center of Advanced Microstructures, Nanjing, 210093, China}

\begin{abstract}
{\bf In order to understand the long-standing problem of the nature of glass states, we performed intensive simulations on the thermodynamic properties and potential energy surface of an ideal glass. We found that the atoms of an ideal glass manifest cooperative diffusion, and show clearly different behavior from the liquid state. By determining the potential energy surface, we demonstrated that the glass state has a flat potential landscape, which is the critical intrinsic feature of ideal glasses. When this potential region is accessible through any thermal or kinetic process, the glass state can be formed and a glass transition will occur, regardless of any special structural character. With this  picture, the glass transition can be interpreted by the emergence of configurational entropies, as a consequence of flat potential landscapes}
\end{abstract}

\maketitle
The nature of glassy states (GS) is one of the most compelling puzzles and questions facing scientists today.\cite{sciences125} The significance of GS not only lies in the importance of their various applications, but also stems from the challenge to current theories in condensed matter physics and thermodynamics.\cite{Angell1995,Debenedetti2001} Until now, it is probably safe to say that, both GS and the glass transition (GT) are within a dark area in the field of materials science and condensed matter physics. Given that a glass is neither a normal liquid nor a standard solid, GS are quite often not described in any detail by standard textbooks.\cite{Berthier2011}  Although the problems have been recognized for over a century and great efforts have been made on various aspects of GS and GT, from disordered structures to kinetic fragility, from dynamic heterogeneity to thermodynamic models,\cite{Berthier2011,Dyre2006,Cavagna2009,Wang2012,Matthieu2016} the debate on essential issues is still quite lively.

One of most crucial issues regarding  GS is that below the Kauzmann temperature $T_K$  the entropy of a supercooled liquid by extrapolation becomes lower than the crystal entropy.\cite{Kauzmann1948} This constitutes the so-called Kauzmann paradox, which thermodynamic models face. To avoid the paradox, an unusual state of matter, namely an ideal glass, was proposed.\cite{DiMarzio1958, Turnbull1959, Angell1968, Wolynes1989} It is defined as a glass state with zero configurational entropy. For more than half a century, the ideal glass remained a conjecture, as its existence has been doubtful in real materials.\cite{Biroli2012} In 1998, two of the present authors (DY and XG)  demonstrated that an $Al_{43}$ cluster in its most stable configuration shows a typical glass transition character,\cite{Gong1998} which can be identified as an ideal glass.\cite{Calvo2006} Ten years later, the ideal glass was found experimentally in that system.\cite{Jarrold2007} To our  knowledge, the $Al_{43}$ cluster  may be the first ideal glass theoretically predicted in real materials. It thus provides an ideal system to study the nature of the GS and GT at the atomic level. In this letter, we report the results of molecular dynamics (MD) simulations and stochastic surface walking (SSW)\cite{shang2013,zhang2013} studies of the thermodynamic properties of the ideal glass $Al_{43}$. For comparison, the symmetric $Al_{55}$ with $I_h$ symmetry and $Al_{40}$ with $D_{6h}$ symmetry were also studied.\cite{Doye2003} The SSW algorithm was recently developed by Shang and Liu and has been proven to be an efficient tool to explore the potential landscape of both clusters and periodic systems.\cite{shang2015,wei2015} An automated climbing mechanism is implemented in the SSW algorithm to manipulate a structural configuration from a local minimum to a high-energy configuration along one random mode direction. We found that the intrinsic nature of an ideal glass is closely associated with a flat potential landscape.

We adopted the $glue$ potential to described the atomic interaction of Al. This potential can correctly reproduce many basic properties of aluminum in crystalline and non-crystalline phases.\cite{Ercolessi1994} The constant-temperature MD method without any boundary conditions was used in the calculations, and any global motion of clusters was carefully eliminated. To obtain a reliable result, the simulation spanned the sub-microsecond scale in sub-picosecond steps. To characterize the shape of a cluster, the principal radii of gyration were calculated. In the present calculations, the heat capacity was based on a constant-temperature ensemble, and the volume of clusters was calculated based on the Wigner-Seitz primitive cell approximation.\cite{Sun2002} Diffusion constants, which are an important thermodynamic quantity for identifying phases, were obtained by calculating the mean-square displacement of each atom. To characterize a cooperative motion among atoms, we define the time-dependent displacement ($R_c(t)$) as
 \[R_c(t)=\frac{1}{N}\sum_i^N [r_i(t)-r_i(t+\Delta t)]^2\]
 with $\Delta t=5$ ps.

 Here we also provided the direct potential landscape from an SSW calculation. For each cluster, we  performed 10 parallel SSW runs and collected 105 minima. To identify cluster structures, a distance-weighted Steinhardt-type order parameter\cite{steinhardt1983} was used to distinguish distinct minima.

Fig. 1 shows the current calculations of the energy, volume and specific heat of $Al_{43}$. One can see that the continuous change in the energy (upper panel), volume (middle panel), and specific heat (lower panel), indicates typical features of the GT. Similar calculations were also performed on $Al_{55}$ and $Al_{40}$ with the results showing that both exhibit typical first-order transition behavior,  in agreement with previous studies.\cite{Gong1998} From Fig. 1, one can see that around 350 K, referred to as the starting temperature of the GT ($T_s$) (see below), the specific heat has a small jump. Meanwhile the shape of clusters begins to change, as indicated by the principal radii of gyration (right panel of Fig. 1).   Since the system is in a well-defined liquid state above 550 K (see below), the GT range can be defined from 350 K to 550 K. In the following we will focus on this temperature range.

We found that, before melting to a normal liquid state, atoms in GS display a remarkable cooperative diffusion, which is intrinsically different from that exhibited by liquid states. Fig. 2 depicts the diffusion constants versus temperature for the three clusters. It can be seen that above 550 K the diffusion constants show a clear Arrhenius temperature dependence, and all three clusters are in the same liquid states with the same diffusion constants. At low temperatures, before melting, no diffusion was detected for $Al_{55}$ and $Al_{40}$. However, we clearly identified  an Arrhenius temperature dependence for diffusion in the GS for $Al_{43}$ as shown in Fig. 2 (see the black dashed line). This was quite surprising, since for conventional GS or supercooled liquids, diffusive behavior is usually believed to have a non-Arrhenius character.\cite{Dyre2006, Gibbs1965} More surprising, contrary to the intuitive picture, the activation energy in the GS is even smaller than that in the liquid state. This shows that the GS and the liquid state are different states. This is strong evidence that the GS is not a supercooled liquid state.

All the atoms contribute to the diffusion in the GT range, not only surface atoms. To illustrate this point, we calculated the diffusion constant of each atom versus its average position, which is shown in the lower panel of Fig. 3. At 600 K, the cluster is in a well-defined liquid state. As expected, the diffusion constant is independent of the atomic position. While at 460 K, even inner atoms in the cluster have significant diffusion, similar to the surface atoms. At this temperature the cluster is still in a solid state, and the solid-solid structural transition is a straightforward explanation of the atomic diffusions.

To further confirm the solid-solid phase transition in the range of the GT, we calculated $R_c(t)$ at three different temperatures as shown in the upper panel of Fig. 3. At 200 K, $Al_{43}$ is obviously in a well-defined solid state, the atoms vibrate around their equilibrium positions, and no diffusion occurs. In this case, $R_c(t)$ has a very small amplitude, and a small variation with time. At 600 K in the liquid state, although $R_c(t)$ is large, it is consistent with time. This is typical behavior for atomic diffusion of liquids due to less cooperation among atoms. On the other hand, at 460 K in the GS, $R_c(t)$ shows a remarkable variation with time, with some periods essentially as for 200 K and others consistent with 600 K. This indicates that atom diffusion is induced by a cooperative motion in a short time, $i.e.$, a solid-solid transition. It should be noted that the cooperative motion in ideal GS is different from that reported in supercooled molecular liquids, which display a non-Arrhenius character.\cite{Dyre2006, Gibbs1965}

The above results suggest that the GT being associated with solid-solid structure transitions implied that the GS must be in a region with flat potential landscape. The phase diagrams for $Al_{43}$ and $Al_{55}$ based on those minima are shown in the upper panel of Fig. 4. Possible phase transformation paths are shown in the lower panel of Fig. 4, where each step had the lowest barrier to the most accessible metastable configuration from the current local minima. As expected, the symmetric cluster ($Al_{55}$) has high stability, reflected in a larger energy difference (0.4 eV) between the ground state and the second most stable state, as well as a large energy barrier of about 0.75 eV from the ground state to the most accessible metastable state. On the contrary, for the ideal glass $Al_{43}$, the energy difference between the ground state and the second most stable state is very small, only 0.06 eV. The energy barrier from the ground state to the most accessible metastable state is also small, only 0.27 eV. This energy barrier should be responsible for the onset of the GS, namely $T_s$. Once the cluster escapes from the ground state, the energy barriers among different states become much smaller (in the order of 0.1 eV), and the GT begins. Here $T_s$ can be considered as the Kauzmann temperature ($T_K$),\cite{Kauzmann1948} below which the configurational entropy of an ideal glass becomes zero.

From  the above results and discussions, we are in a position to present a general thermodynamic picture for ideal glasses, which is schematically plotted in Fig. 5. A structural phase transition driven by temperature usually involves an entropy increase in the high-temperature phase. As previously suggested,\cite{Johari2008} we neglected the difference of vibrational entropies in solid and liquid states, we focused only on the configurational entropy. Below $T_s$ ($T_K$), since ideal GS has zero configurational entropy, it should behave as a conventional solid (red-dot-dash line in Fig. 5). However above $T_s$ the configurational entropy begins to emerge, and the free energy of an ideal glass becomes lower (blue solid line in Fig. 5). With the increase of temperature, the ideal glass can visit more and more configurations, consequently the configurational entropy keeps increasing until the melting temperature ($T_m$ in Fig. 5) is reached. Here the free energy curves show a crossover between the ideal GS and liquid state, because the ideal glass is a distinct state rather than a frozen-in supercooled liquid. On the contrary, for GS with a frozen-in supercooled liquid,\cite{Goldstein1969,Debenedetti2001,Parisi2002} the free energy curves will merge into that of liquid state near $T_m$, rather than a crossover.

In summary, we have demonstrated that, an ideal glass state is a distinct state rather than  a frozen-in supercooled liquid. The intrinsic feature of an ideal glass is a flat potential landscape. The glass transition of an ideal glass can be viewed as the process of gaining more and more configurational entropy. We believed that the glass transition only depends on the special potential landscape not the structural details.
\appendix*

\section*{Acknowledgements}

This research is supported by the Natural Science Foundation of China, National Basic Research Program of China. The computation is performed in the Supercomputer of East China Normal University.

\newpage

\begin{figure}[figure1]
\centering
\includegraphics[width=70mm]{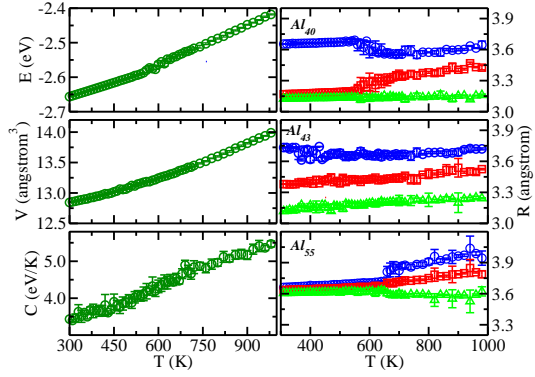}
\caption{(Color online) Left panel: the energy (upper), volume (middle) and specific heat (lower) as a function of temperature for $Al_{43}$. The continuous change shows a typical glass transition behavior. Right panel: three principal radii of gyration ($R_{1,2,3}$ indicated by different colors) for $Al_{40}$ (upper), $Al_{43}$ (middle), and $Al_{55}$ (lower). Around $T_s$ ($\sim$350 K), the shape of $Al_{43}$ begins to change, while the shape of $Al_{40}$ and $Al_{55}$ is unchanged until melting.}
\end{figure}

\begin{figure}[figure2]
\centering
\includegraphics[width=70mm]{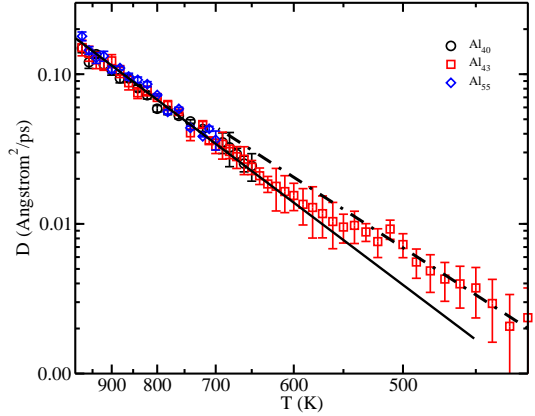}
\caption{(Color online) Diffusion constants versus temperature for $Al_{40}$, $Al_{43}$ and $Al_{55}$, where the y-axis is logarithmic and the x-axis is reciprocal temperature. The dashed and solid lines are the Arrhenius fits to the data. The difference in slope of the fitted lines indicates different activation energies.}
\end{figure}

\begin{figure}[figure3]
\centering
\includegraphics[width=70mm]{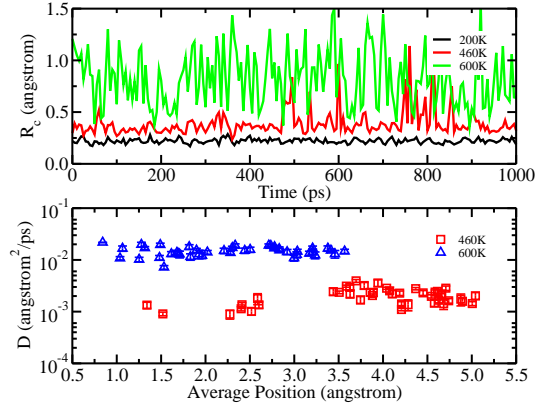}
\caption{(Color online)
Upper panel: the time-dependent displacement for three temperatures. At 460 K, although the displacement is notable, it only occurs in several short durations, which indicates a cooperative motion of atoms in the glass transition range. Lower panel: the diffusion constant of each atom versus its average position. In the glass transition range, the inner atoms have non-negligible diffusion. }
\end{figure}

\begin{figure}[figure4]
\centering
\includegraphics[width=70mm,angle=0]{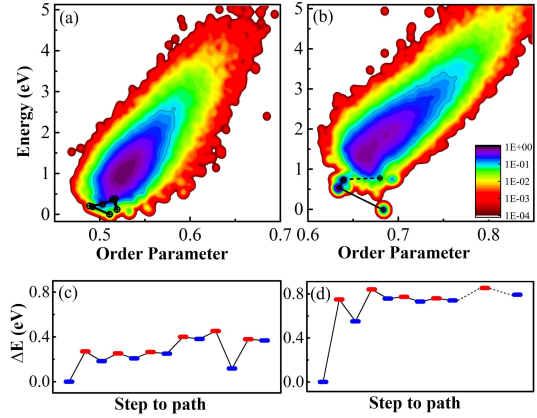}
\caption{(Color online) Upper panel: the phase diagram of $Al_{43}$ (left) and $Al_{55}$ (right). The black line shows the minima along a possible path for configuration changing from the ground state. Lower panel: the energy profile of the path indicated in upper panel. The blue bars correspond to the minima while the orange bars correspond to the transition states.}
\end{figure}

\begin{figure}[figure5]
\centering
\includegraphics[width=70mm]{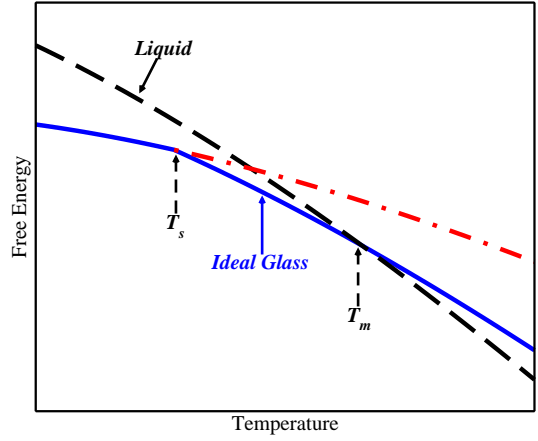}
\caption{(Color online) The schematic phase diagram for an ideal glass (blue solid line) and a liquid (black dashed line). $T_m$ indicates the temperature at which the free energies of the liquid and ideal glass states are equal. $T_s$ refers to the emergence of configuration entropies at this temperature. Without these configuration entropies, the free energy of an ideal glass should go along the red dot-dash line for $T>T_s$. }
\end{figure}

\end{document}